\documentclass[twocolumn]{aastex62}

\newcommand{\kms}{km s$^{-1}$$\,$}
\graphicspath{{./}{figures/}}

\received{---}
\revised{---}
\accepted{---}
\submitjournal{ApJ}

\shorttitle{Interstellar Plunging Waves}
\shortauthors{Cosentino et al.}

\begin{document}

\title{Interstellar Plunging Waves: ALMA resolves the physical structure of non-stationary MHD shocks}

\correspondingauthor{Giuliana Cosentino}
\email{giuliana.cosentino.15@ucl.ac.uk}

\author{Giuliana Cosentino}
\affil{Department of Physics and Astronomy, University College London, Gower St, London, WC1E6BT, UK}
\affiliation{European Southern Observatory, Karl-Schwarzschild-Strasse 2, 85748 
Garching, Germany}

\author{Izaskun Jim\'enez-Serra}
\affiliation{Departamento de Astrof\'{\i}sica, Centro de Astrobiologia, 28850 Torrej\'on de Ardoz, Madrid, Spain}
\affiliation{School of Physics \& Astronomy, Queen Mary University of London, Mile End Road, E1 4NS, London, UK}

\author{Paola Caselli}
\affiliation{Max-Planck Institute for Extraterrestrial Physics, Gie\ss enbachstrasse 2, 85748 Garching, Germany}

\author{Jonathan D. Henshaw}
\affiliation{Max-Planck Instiutute for Astronomy, K\'onigstuhl 17, 69117 Heidelberg, Germany}

\author{Ashley T. Barnes}
\affiliation{Argelander-Institut f\"ur Astronomie
Auf dem H\"ugel 71, D-53121 Bonn, Germany}

\author{Jonathan C. Tan}
\affiliation{Space, Earth and Environment Department, Chalmers University of Technology, Chalmersplatsen 4, 41296 G\"oteborg, Sweden }
\affiliation{Department of Astronomy, University of Virginia, Charlottesville, Virginia, USA}

\author{Serena Viti}
\affiliation{Department of Physics and Astronomy, University College London, Gower St, London, WC1E6BT, UK}

\author{Francesco Fontani}
\affiliation{INAF-Osservatorio Astrofisico di Arcetri, Largo E. Fermi 2, 
I-50125 Firenze, Italy}

\author{Benjamin Wu}
\affiliation{National Astronomical Observatory of Japan, Yubinbango 181-8588 Tokio, Mitaka, Osawa 2-21-1, Japan}



\begin{abstract}
Magneto-hydrodynamic (MHD) shocks are violent events that inject large amounts of energy in the interstellar medium (ISM) dramatically modifying its physical properties and chemical composition. Indirect evidence for the presence of such shocks has been reported from the especial chemistry detected toward a variety of astrophysical shocked environments. However, the internal physical structure of these shocks remains unresolved since their expected spatial scales are too small to be measured with current instrumentation. Here we report the first detection of a fully spatially resolved, MHD shock toward the Infrared Dark Cloud (IRDC) G034.77-00.55. The shock, probed by Silicon Monoxide (SiO) and observed with the \textit{Atacama Large Millimetre/sub-millimetre Array} (ALMA), is associated with the collision between the dense molecular gas of the cloud and a molecular gas flow pushed toward the IRDC by the nearby supernova remnant (SNR) W44. The interaction is occurring on sub-parsec spatial scales thanks to the enhanced magnetic field of the SNR, making the dissipation region of the MHD shock large enough to be resolved with ALMA. Our observations suggest that molecular flow-flow collisions can be triggered by stellar feedback, inducing shocked molecular gas densities compatible with those required for massive star formation. 

\end{abstract}



\section{Introduction} \label{sec:intro}
Interstellar MHD shocks are known to exist in a wide variety of astrophysical environments such as jets and molecular outflows \citep{martinpintado1992,caselli1997,girart2001}, stellar winds \citep{wilkin1996,babel1997} and supernovae explosions \citep{gotthelf2001,grefenstette2014,miceli2019}. Depending on the propagation speed, pre-shock gas density and magnetic field strength, MHD shocks can be either J-type (from \textit{jump}, fast shocks with $\rm v_s$$\geq$45-50$\,$\kms and with a discontinuity in the gas physical properties) or C-type \cite[more gentle interactions found in molecular clouds with low ionization fractions, where the gas physical properties change in a \textit{continuous} manner;][]{draine1980,draine1983,neufeld1989}. If the MHD shock has not attained the steady state, theory predicts the presence of a J-type discontinuity within the structure of the shock at the early stages of its propagation (the {\it piston}), which later on evolves into a C-type shock \citep[a non-stationary CJ-type shock;][]{chieze1998,flower2003,lesaffre2004}.
 
In star forming regions, MHD shocks are expected to develop across spatial scales of $\sim$10$^{14}$-10$^{15}\,$cm \citep[$\sim$0.05-0.5$"$ at a distance of 140 pc;][]{gusdorf2008}, too small to be easily resolved with current instrumentation. However, interstellar shocks are also predicted to arise when large-scale flows of molecular gas are pushed to collide. In such collisions, the shock occurs across parsec/sub-parsec spatial scales ($\sim$3$\times$10$^{17}$-3$\times$10$^{18}\,$cm), sufficient to spatially resolve its physical and dynamical evolution \citep{wu2015,inutsuka2015}.

In this Letter, we report the first images of the internal physical structure of a non-stationary CJ-type shock. The emission of the typical shock tracer SiO has been mapped with ALMA toward the IRDC G034.77-00.55 (hereafter G034), which is located at 2.9$\,$kpc in a highly dynamical environment between the H{\small II} region G034.8-00.7 and the supernova remnant (SNR) W44. The interaction between G034 and W44, located at the same distance, has long been established  \citep[e.g.][]{seta1998,seta2004,claussen1997,ortega2007,yoshiike2013,ranasinghe2018}, making G034 an ideal target for MHD shock investigation. Our ALMA images show that a molecular gas flow pushed by the expanding shell of W44 is interacting with a dense molecular ridge in G034 forcing the molecular flow to decelerate and {\it plunge} onto the IRDC. This interaction enhances the density of the gas in the ridge to levels required for the formation of the most massive stars.

\section{Observations} \label{sec:obser}
We used ALMA Band 3 during Cycle 4 (P.I. I. Jimenez-Serra) to map the SiO J=2$\rightarrow$1 transition (86.85$\,$GHz) toward G034. Observations were performed using the ALMA Compact Array (11 antennas) plus 45 antennas of the 12m-array (baseline 15$\,$m-331$\,$m; uv distance 4.3-9.6$\,$k$\lambda$). Observations were perfomed in dual polarization mode, using a spectral bandwidth of 58.594$\,$MHz. The sources J1751$+$0939 and J1851$+$0035 were used as calibrators. We reproduced the two calibrated datasets by running the original pipeline reduction scripts and used the task \textit{uvcontsub} in {\sc CASA} version 4.7.2\footnote{https://casa.nrao.edu/} to subtract the continuum emission previously estimated from line-free channels. The two continuum-subtracted datacubes were combined using the task \textit{concat}. We generated the SiO line-only final image using the task \textit{clean} with Briggs weighting (robust parameter 0.5), phase centre $\alpha$(J2000)=18$^h$56$^m$41.7$^s$, $\delta$(J2000)=1$^{\circ}$23$^{\prime}$25$^{\prime\prime}$ and velocity resolution 0.1$\,$\kms\space ($\sim$30 kHz). The final high-angular resolution images have synthesised beam of 3.5\arcsec$\times$2.5\arcsec, PA= 84.27$^{\circ}$ and noise level 10$\,$mJy$\,$beam$^{-1}$. By comparing the ALMA and IRAM 30$\,$m spectra extracted toward the position of the SiO peak detected by \citet{cosentino2018} in this cloud, we find that no significant flux is missing in the 12m+ACA SiO image. The 3$\,$mm continuum map obtained with the 12m-array does not show any continuum point$-$like source spatially associated with SiO and above the 3$\sigma$ level (1$\sigma$=66$\,\mu$Jy$\,$beam$^{-1}$) . 

Observations of the C$^{18}$O J=1$\rightarrow$0 (109.78 GHz) and J=2$\rightarrow$1 (219.56 GHz) lines were carried out with the IRAM 30m telescope in July 2012. The On-The-Fly (OTF) maps were obtained using the off position (-240\arcsec, -40\arcsec) and a dump speed of 6\arcsec$\,$s$^{-1}$. Pointing accurcacy was $\leq$3\arcsec \space and line calibrations were performed on the source G34.4$+$0.3. The FTS spectrometer (4$\,$GHz full bandwidth) provided a spectral resolution of 200 kHz i.e. 0.54$\,$\kms at 109.78$\,$GHz and 0.27$\,$\kms at 219.56$\,$GHz. The final data cubes, created using the {\sc CLASS} software within {\sc GILDAS}\footnote{https://www.iram.fr/IRAMFR/GILDAS/}, have spatial resolution and beam efficiency of 24\arcsec\space and 0.78 and  for C$^{18}$O(1-0) and 12\arcsec\space and 0.63 for C$^{18}$O(2-1). The final rms is 0.1 K for both spectra.

\section{Results} \label{sec:results}

\subsection{SiO ALMA data: moment maps and Position-Velocity (PV) diagrams}
\label{sio}
\begin{figure}
    \centering
    \includegraphics[scale=0.7]{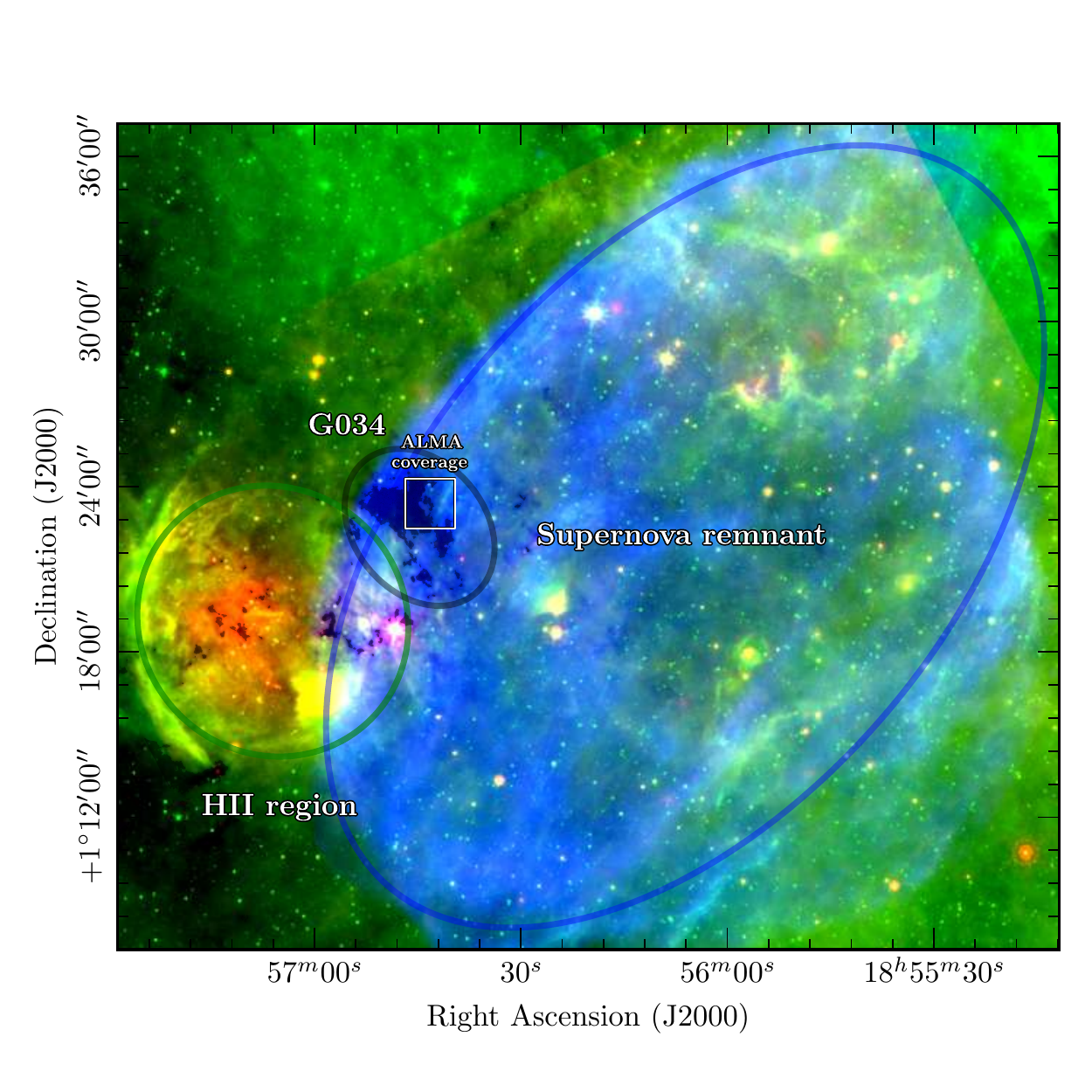}
    \caption{Three-colour image of G034 showing its position (black circle) between W44 (blue circle) and G034.758-00.681 (green circle). Red is 24$\,$$\mu$m emission \citep[Spitzer MIPSGAL;][]{carey2009}, green is 8$\,$$\mu$m emission \citep[Spitzer GLIMPSE;][]{churchwell2009}, and blue shows a combined JVLA+GBT 21cm continuum map \citep[THOR survey;][]{beuther2016}. The white square indicates the ALMA mosaic area and the grey shadow corresponds to A$_v\,\geq$20$\,$mag.}
    \label{fig1}
\end{figure}

In Figure$\,$\ref{fig1}, we present the three-colour image of G034 (black circle) located between W44 (blue circle) and the HII region G034.758-00.681 (green circle). Red is 24$\,$$\mu$m emission \citep[Spitzer MIPSGAL;][]{carey2009}, green is 8$\,$$\mu$m emission \citep[Spitzer GLIMPSE;][]{churchwell2009}, and blue shows a combined Jansky Very Large Array (JVLA) and Green Bank Telescope (GBT) 21cm continuum map \citep[THOR survey;][]{beuther2016}. The white square indicates the extent of the ALMA mosaic. The molecular gas around W44 is expanding at a velocity of $\sim$11-13$\,$\kms \citep{sashida2013}, although in regions close to G034 it shows terminal velocities $\sim$20-25$\,$\kms \citep[see region W44F;][]{anderl2014}. The molecular gas outside W44 moves with $\rm v_{\rm LSR}$$\sim$45-50$\,$\kms \citep[i.e. distances $S$$\geq$1 in Figures$\,$6 and 7 of][]{sashida2013}.   

\begin{figure}
    \centering
    \includegraphics[scale=0.50,trim =3cm 0cm 0cm 0cm, clip=True]{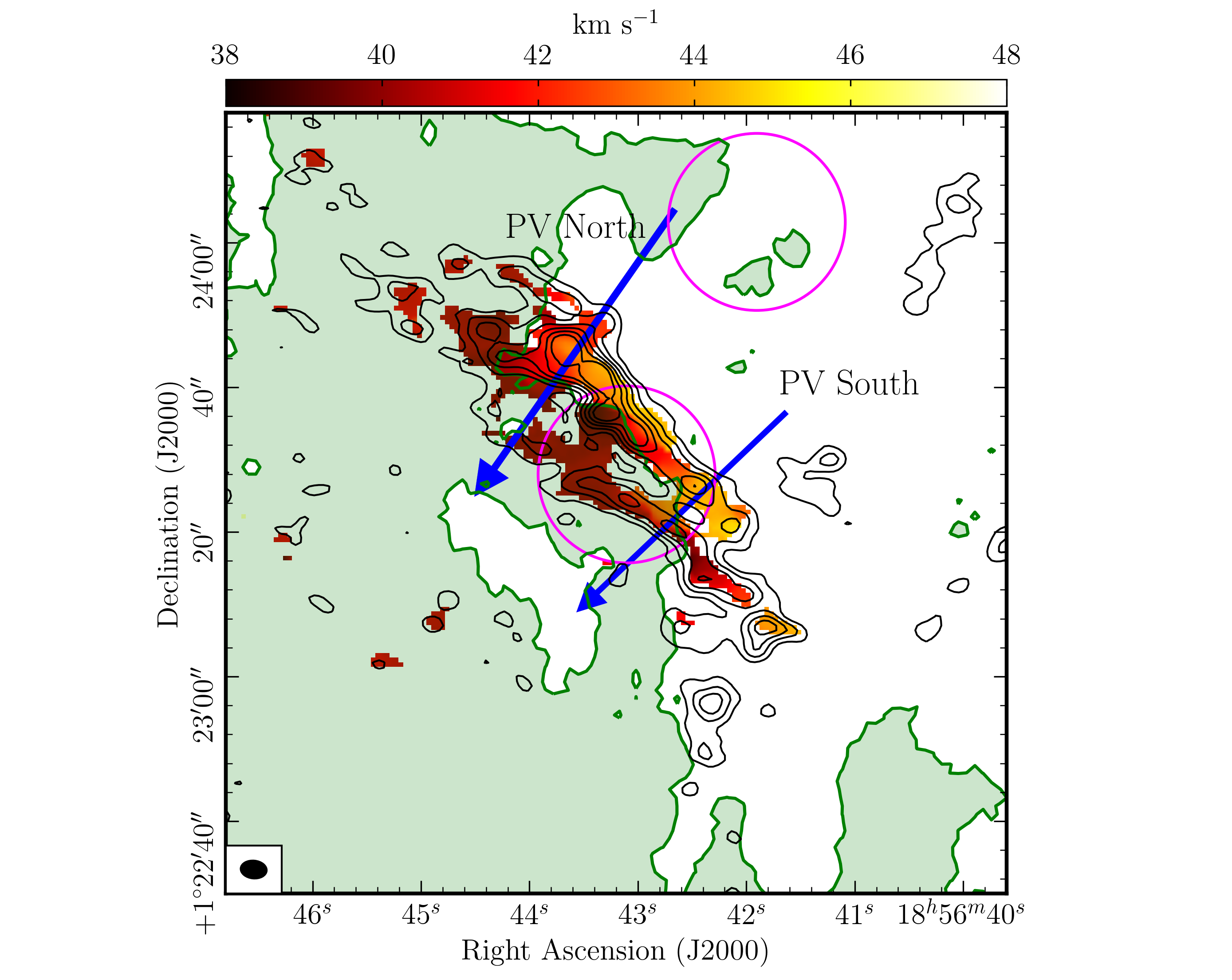}
    \caption{SiO integrated intensity map (black contours; velocity range 36.6$-$47.6$\,$\kms) toward G034 superimposed on its moment 1 velocity map (red scale). Contours are from 3$\sigma$ ($\sigma$= 0.016$\,$Jy$\,$\kms) by steps of 3$\sigma$. Green contour and shadow indicates the A$_v\geq$20$\,$mag dense material in the IRDC \citep[][]{kainulainen2013}. Blue arrows show the directions used to extract the PV diagrams of Figure$\,$\ref{fig4}. Magenta circles indicate the two positions used to estimate the gas density conditions. Beam size is indicated as black ellipse in the bottom left corner.}
    \label{fig2}
\end{figure}

\begin{figure*}
    \centering
    \includegraphics[scale=0.50]{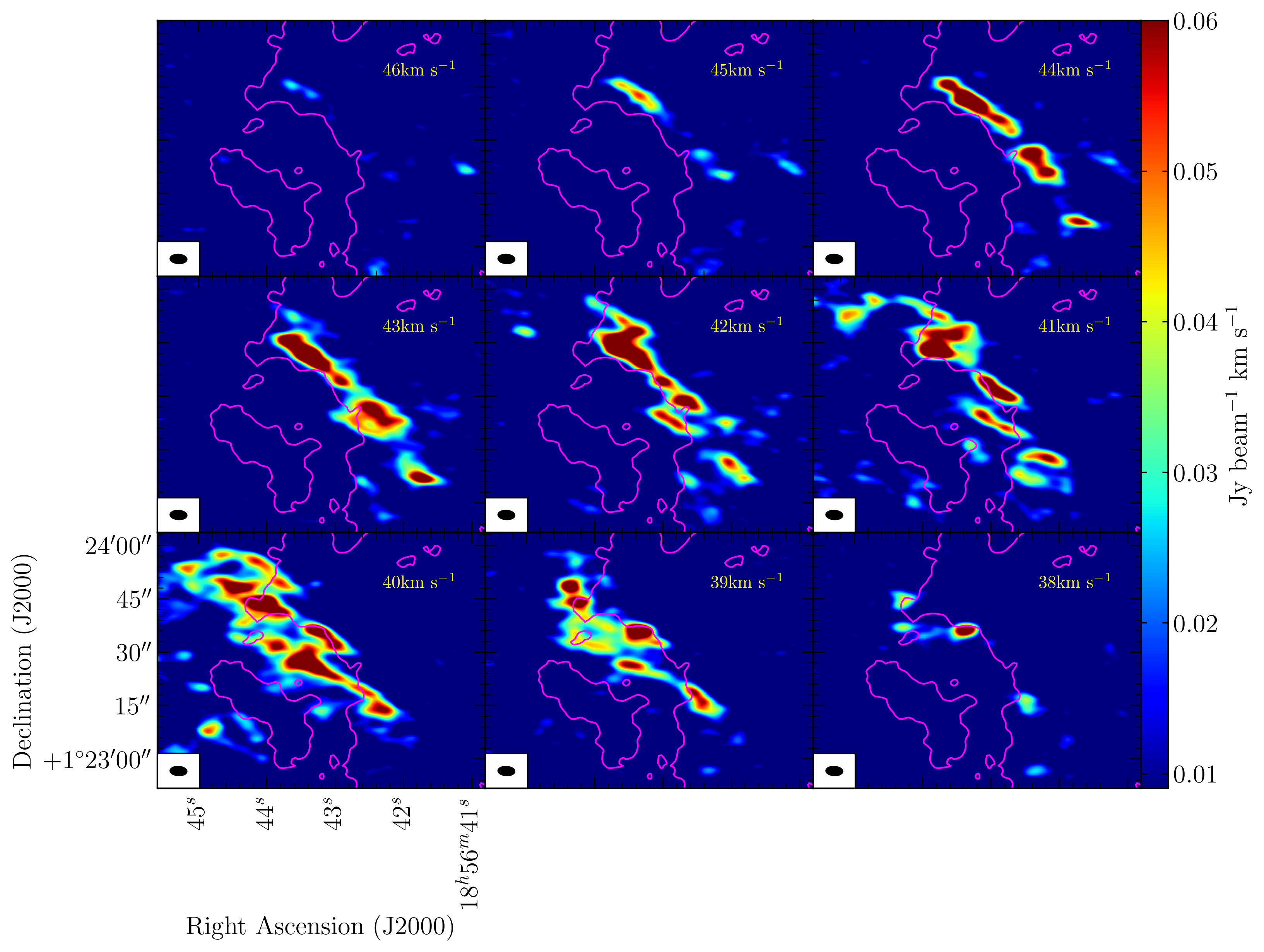}
    \caption{SiO channel maps for the velocity range 38$-$46$\,$km$\,$s$^{-1}$. Integration step is 1$\,$km$\,$s$^{-1}$ and intensities are $\geq$3$\sigma$ ($\sigma$= 0.006$\,$Jy$\,$beam$^{-1}$$\,$km$\,$s$^{-1}$). The magenta contour \citep[with Av$\sim$20$\,$mag][]{kainulainen2013} indicates the position of the high-density ridge found toward the northwest of G034. Beam size and central velocity are shown for each map.}
    \label{fig3}
\end{figure*}

Figure~\ref{fig2} reports the integrated intensity map (black contours; velocity range 36.6$-$47.6 \kms) and the velocity moment 1 map (red scale) of the SiO emission measured toward G034 with ALMA. These maps appear superimposed on the A$_{\rm v}$=20$\,$mag visual extinction level of the cloud \citep[][]{kainulainen2013}, which shows a dense ridge toward the northwest of G034. The SiO emission is organised into two plane-parallel structures: a bright elongation peaking outside the ridge at red-shifted velocities with respect to the radial velocity of the IRDC ($\rm v_{\rm LSR}$=43$\,$\kms); and a fainter elongation found into the ridge between $\sim$36-42$\,$\kms. The SiO channel maps (Figure$\,$\ref{fig3}) reveal that shifting from 46$\,$\kms to 39$\,$\kms, SiO first appears in a region outside the Av$\sim$20$\,$mag ridge, to progressively move toward the cloud bending and spreading onto it (as if it had impacted).

\begin{figure*}
    \centering
    \includegraphics[scale=0.5]{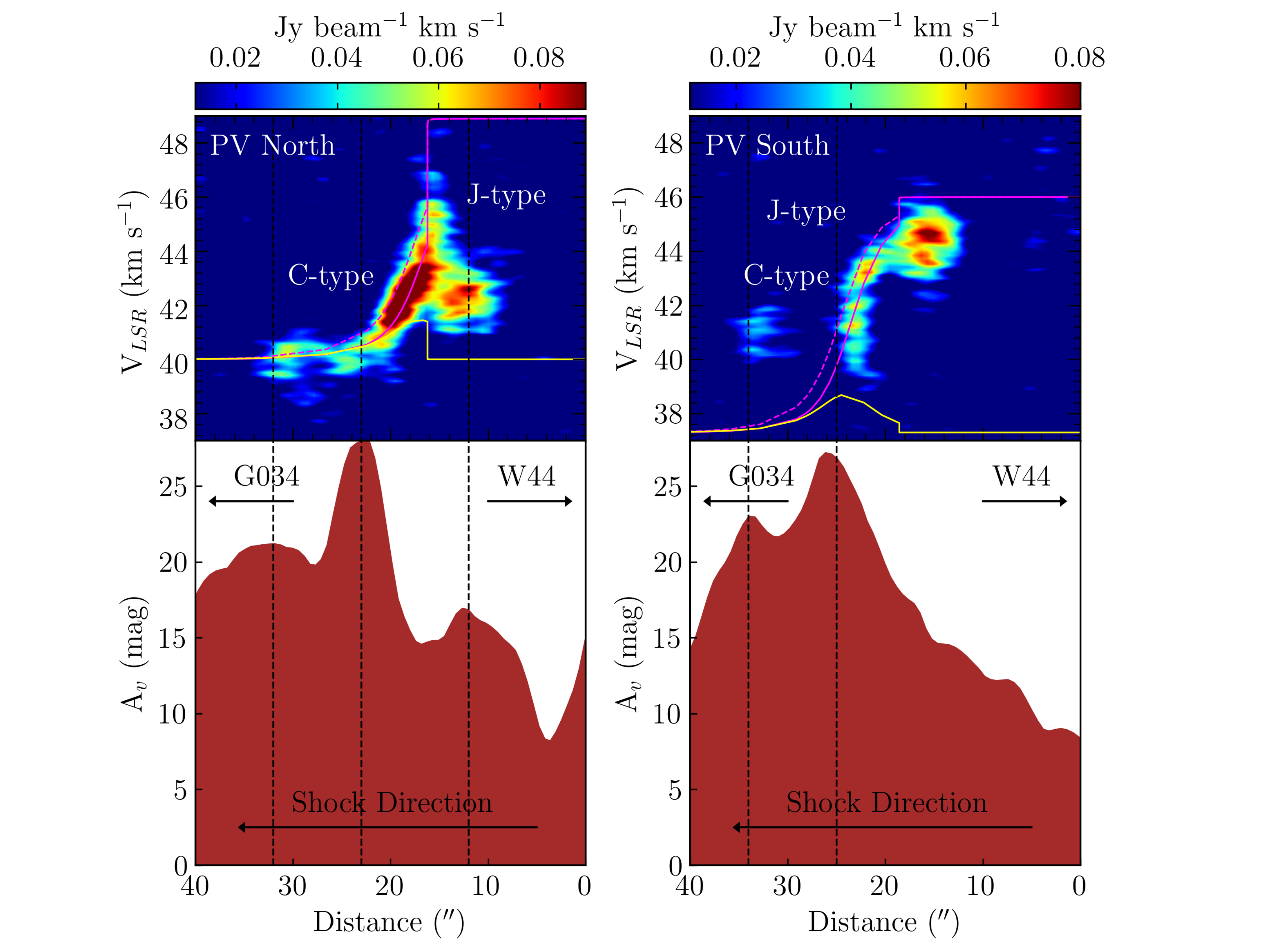}
    \caption{SiO (colour scale) North (top left) and South (top right) PV diagrams and corresponding A$_{v}$ profiles (bottom panels) extracted across the two SiO elongations. Velocity profiles predicted by the MHD$\_$VODE code for the neutral (solid magenta line) and charged (dashed magenta line) fluids are shown along with the drift velocity ($|v_i-v_n|$) between the two fluids (yellow line). Vertical dashed lines (in black) show the local peaks in extinction.}
    \label{fig4}
\end{figure*}

To analyse the velocity structure of the SiO shocked gas, we have extracted PV diagrams along two representative directions perpendicular to the observed SiO elongations, PV north and PV south (blue arrows in Figure~\ref{fig2}). Figure$\,$\ref{fig4} shows that the deceleration of the gas is occurring in two steps: a first almost vertical velocity decrease of $\sim$2-3$\,$\kms within a spatial width lower than the angular resolution ($\leq$2.5\arcsec), and a second and shallower deceleration of $\sim$4-5$\,$\kms  measured across $\sim$10-20\arcsec. This resembles the predicted velocity structure of CJ-shocks \citep[Section$\,$\ref{model} and][]{chieze1998,lesaffre2004}. The comparison between the SiO PV diagrams and the visual extinction profiles of G034 along the PV directions, reveals that the observed SiO spatially coincides with extinction peaks across the ridge (bottom brown curves in Figure~\ref{fig4}). This is consistent with the idea that the two shocked plane-parallel structures (probed by SiO) appear once the molecular gas expanding from W44 encounters the {\it high-density wall} of the ridge. 

\subsection{C$^{18}$O Single-dish Spectra: H$_2$ Density and SiO Abundance Enhancement in the Post-shock Gas}
\label{c18o}

\begin{figure}
    \centering
    \includegraphics[scale=0.5]{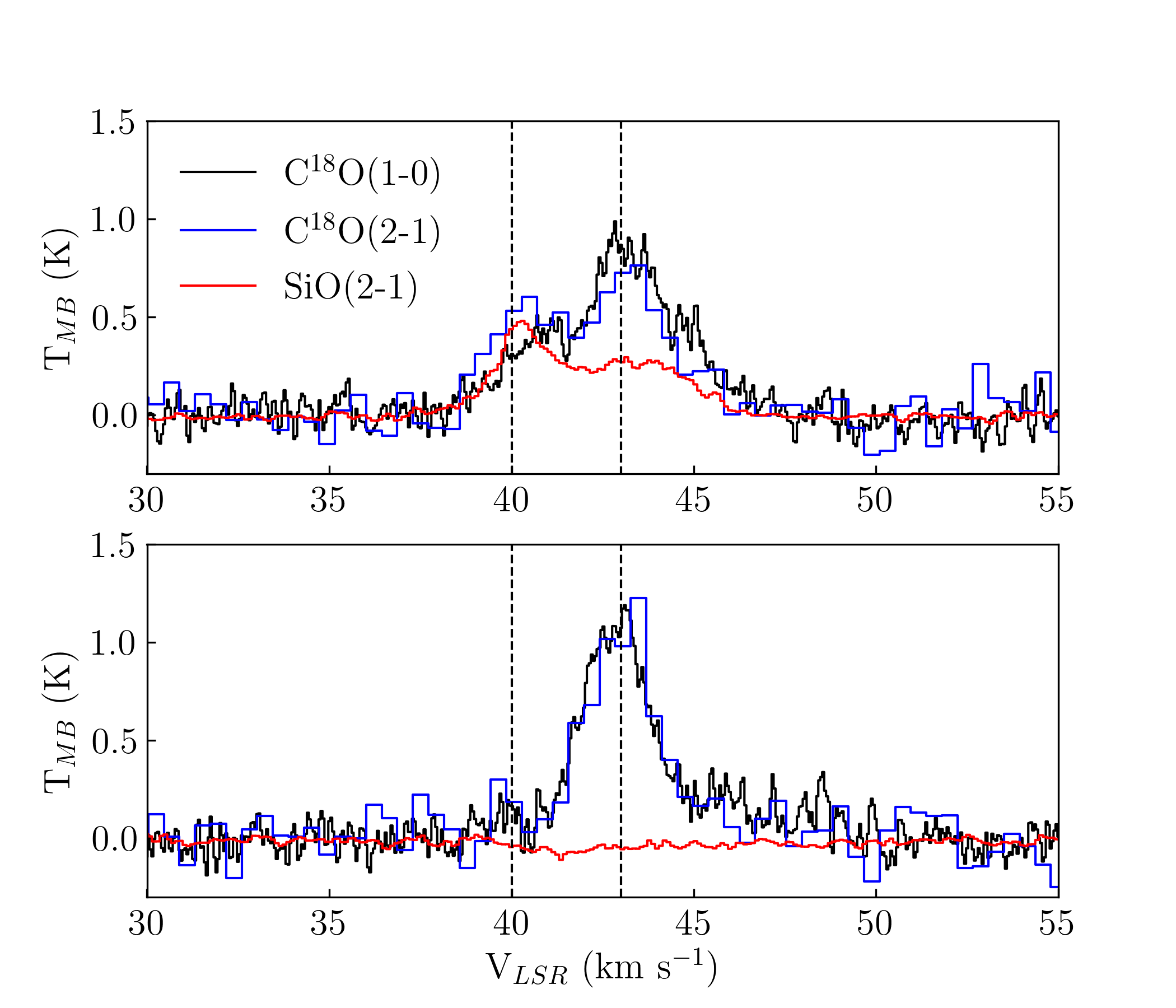}
    \caption{Spectra of C$^{18}$O(1-0) (black lines), C$^{18}$O(2-1) (blue lines) and SiO(2-1) (red lines) extracted toward G034 at positions $\alpha$(J2000)=18$^h$56$^m$43.1$^s$, $\delta$(J2000)=1$^{\circ}$23$^{\prime}$28\arcsec (top panel) and $\alpha$(J2000)=18$^h$56$^m$41.9$^s$, $\delta$(J2000)=1$^{\circ}$24$^{\prime}$03\arcsec over a 25\arcsec-beam. The components at 43$\,$\kms  and 40$\,$\kms corresponds to pre and post-shock gas, respectively (vertical dashed lines).}
    \label{fig5}
\end{figure}

Figure~\ref{fig5} shows spectra of C$^{18}$O(1-0) (black) and C$^{18}$O(2-1) (blue) measured over a 25\arcsec-beam toward two positions in G034, one outside and one inside the ridge (magenta circles in Figure$\,$\ref{fig2}). Outside the ridge, where no SiO is detected, only the quiescent (pre-shock) component at $\sim$43$\,$\kms can be seen (Figure$\,$\ref{fig2}, bottom panel). However, inside the ridge, a second component at $\sim$40$\,$\kms clearly appears coinciding with the detection of SiO (the post-shock gas; upper panel).

By using the radiative transfer code RADEX \citep{tak2007}, we can infer the H$_2$ gas volume density of the pre ($\sim$43$\,$\kms) and post-shock gas ($\sim$40$\,$\kms) in G034 from the C$^{18}$O J=1$\rightarrow$0 and J=2$\rightarrow$1 emission. We assume kinetic temperature of 15$\,$K \citep[typical of IRDCs;][]{pillai2007} and mean linewidth of 2$\,$\kms. The  physical properties of the pre-shock gas at 43$\,$\kms are n(H$_2$)$\sim$10$^4$$\,$cm$^{-3}$, N(C$^{18}$O)$\sim$1.8$\times$10$^{15}\,$cm$^{-2}$, $\tau\sim$0.17 and Tex$\sim$9$\,$K, while for the post-shock gas at 40$\,$\kms the inferred values are n(H$_2$)$>$10$^5$$\,$cm$^{-3}$, N(C$^{18}$O)$\sim$9$\times$10$^{14}$$\,$cm$^{-2}$, $\tau\sim$0.36 and T$_{ex}\sim$7$\,$K. Hence, the density of the post-shock gas is a factor of $\geq$10 higher than the density of the pre-shock gas. Note that the estimated excitation temperatures are comparable to those reported in \citet{cosentino2018} for CH$_3$OH, another shock tracer. 

We also estimate the SiO total column density, N(SiO), for the pre-shock ($\sim$43 \kms) and post-shock gas ($\sim$40 \kms). Since no SiO emission is present outside the ridge, the 3$\sigma$ upper limit for N(SiO) in the pre-shock gas is 6$\times$10$^9$$\,$cm$^{-2}$ by considering rms noise level of 10$\,$mJy$\,$beam$^{-1}$. For the post-shock gas, the derived N(SiO) is 1.6$\times$10$^{12}$$\,$cm$^{-2}$. Hence, the abundance of SiO has been enhanced by a factor $\geq$500 in the post-shock gas in G034, as inferred from the N(SiO)/N(C$^{18}$O) column density ratios. 

\subsection{3$\,$mm Continuum Emission}
\label{3mm}

From the 3$\sigma$ rms level measured in our ALMA 3$\,$mm continuum image (0.2 mJy), we can estimate an upper limit to the envelope mass of any possible proto-star present in the region as:

\begin{equation}
 M_{env} = \frac{F_{\nu} d^2}{B_{\nu}(T_{dust}) k_{\nu}} \frac{M_{gas}}{M_{dust}},
\end{equation}

\noindent
where F$_\nu$ is the continuum flux, $d$ is the source distance, k$_\nu$ is the dust opacity, B$_\nu$(T$_{dust}$) is the Planck Function at the dust temperature T$_{dust}$, and M$_{gas}$/M$_{dust}$ is the gas-to-dust mass ratio \citep{hennings1994}. We assume $d$=2.9$\,$kpc, T$_{dust}$=15$\,$K, M$_{gas}$/M$_{dust}$=100, and F$_\nu$=0.2$\,$mJy. Dust opacity at 3$\,$mm is extrapolated to k$_\nu$=0.0237$\,$m$^2$$\,$kg$^{-1}$ \citep{hennings1994}. From all this, we derive a proto-stellar envelope mass $\leq$0.45$\,$M$_{\odot}$.

\section{Modelling the SiO PV diagrams}
\label{model}

We have used the 1D MHD shock code MHD$\_$VODE \citep{flowerpineau2015} to reproduce the SiO velocity structure measured along the North and South PV diagrams in G034, and to constrain the physical parameters of the shock. We have run a grid of models with shock speeds between $v_s$=5-30$\,$\kms, in accordance with the expanding and shocked velocities of the molecular gas in W44 \citep{sashida2013, anderl2015}. We assume H pre-shock volume densities of n(H)=10$^4$-10$^5$$\,$cm$^{-3}$ (see Section$\,$\ref{c18o}) and consider cosmic-ray ionisation rates $\zeta$=1$\times$10$^{-17}-$5$\times$10$^{-15}$$\,$s$^{-1}$ \citep{vaupre2015}, magnetic field strengths B$_{mag}$=100-1000$\,$$\mu$G \citep{hoffman2005,cardillo2014} and a pre-shock gas temperature of 15$\,$K \citep{pillai2007,ragan2011}. The $\rm v_{\rm LSR}$ of the pre-shock gas ranges from 45-50$\,$\kms, in agreement with the velocities measured outside W44 \citep[Section$\,$\ref{sio} and][]{sashida2013}.

As shown in Figure~\ref{fig4}, the SiO PV diagrams are best reproduced by time-dependent CJ-type shocks that have not reached steady-state yet and that are still evolving. The velocity profiles of the neutral and ion fluids ($v_n$ and $v_i$) are shown by solid and dashed magenta lines respectively, while the drift velocity, $|v_i-v_n|$, is shown by solid yellow lines. PV North can be fitted by a model with n(H)=10$^{4}$$\,$cm$^{-3}$, $v_s$=22.5$\,$\kms, $\zeta$=5$\times$10$^{-16}$$\,$s$^{-1}$, B$_{mag}$=850$\,$$\mu$G and t$_{dyn}$=16000$\,$years. The SiO emission appears at distances where the drift velocity $|v_i-v_n|$$>$0, consistent with the idea that SiO is released by dust erosion/destruction in the shock. However, the SiO column density measured in G034 (1.6$\times$10$^{12}$ cm$^{-2}$; Section$\,$\ref{c18o}) cannot be predicted by the models. This may be due to the fact that the public version of the MHD$\_$VODE code does not include grain-grain collision processes such as shattering. While the threshold drift velocities for the production of SiO by sputtering are $>$20$\,$\kms$\,$\citep{jimenezserra2008,gusdorf2008}, shattering only requires velocities as low as 1$\,$\kms  \citep{guillet2011}. The MHD$\_$VODE code does not consider a small fraction of SiO in the dust mantles either, which may also alleviate the discrepancy between the observed and predicted SiO column densities \citep[see e.g.][]{schilke1997,jimenezserra2008} 

As expected from the interaction with a SNR \citep{hoffman2005,cardillo2014}, the inferred value of B$_{mag}$ in PV North is higher than those measured in molecular clouds with n(H$_2$)$\sim$10$^4$-10$^5$$\,$cm$^{-3}$ \citep[][]{crutcher2010}. In fact, only models with such large B$_{mag}$ values can predict long dissipation lengths as observed in PV North. 

In contrast to PV North, no model can reproduce well the sharp decrease in velocity observed  in PV South. The best-fit model gives n(H)=10$^4$$\,$cm$^{-3}$, $v_s$=23$\,$\kms, $\zeta$=1$\times$10$^{-15}$$\,$s$^{-1}$, B$_{mag}$=880$\,$$\mu$G and 17000$\,$years. This mismatch between the model and the observations could be due to SiO (a high-density tracer) not being excited enough along the whole length of the shock. Alternatively, MHD shock theory may need to be revisited. 

\section{Discussion}
\label{sec: disc}

\subsection{SiO emission caused by protostars and outflows?}
\label{protostars}

Our ALMA 3$\,$mm continuum map does not reveal the presence of any point-like source in G034 associated with the SiO emission down to a proto-stellar envelope mass 0.45$\,$M$_{\odot}$ (Section$\,$\ref{3mm}). This value is comparable to the lower limits of Class 0 and I envelope masses ($>$0.5$\,$M$_{\odot}$; \cite{arce2006}) and well below the typical values for high-mass protostars ($>$100$\,$M$_{\odot}$; \cite{zhang2005}). Therefore, if present, both low-mass and high-mass protostars should have been detected in our 3$\,$mm ALMA image.  

In addition, following \citet{dierickx2015}, we have estimated the mass (M), linear momentum (P) and kinetic energy (E) of the SiO emission in G034 to be M=12.2$\,$M$_{\odot}$, P$\,$=$\,$18.8$\,$M$_{\odot}$$\,$\kms  and E$\,$=$\,$3.8$\times$10$^{41}$$\,$ergs. Outflows driven by low-mass protostars have typical parameters in the range M$\sim$0.005$-$0.15$\,$M$_{\odot}$, P$\sim$0.004-0.12$\,$M$_{\odot}$$\,$\kms  and E $\sim$2$\times$10$^{40}$-5$\times$10$^{42}$$\,$ergs \citep{arce2006}. Although the kinetic energy inferred from SiO is comparable to that in outflows driven by low-mass protostars, the mass and linear momentum are factors 100$-$3000 higher. Outflows driven by high-mass protostars have masses of few tens of M$_{\odot}$ and linear momentum $\sim$10-100$\,$M$_{\odot}$$\,$\kms,  comparable to our values \citep{zhang2005}; but their kinetic energies are 5 orders of magnitude higher than measured in G034 \citep[10$^{46}$$\,$ergs vs. 3.8$\times$10$^{41}$$\,$ergs;][]{zhang2005}). It is thus unlikely that SiO in G034 is associated with star formation activity.

\subsection{A molecular flow-flow collision unveiled by plunging waves}
\label{collision}

G034 presents little star formation activity \citep{rathborne2006,chambers2009} and hence, internal stellar feedback is not expected to affect the pristine physical and chemical conditions of the cloud. In such a quiescent region, molecular shock tracers such as SiO are heavily depleted onto dust grains \citep{martinpintado1992,schilke1997}. \citet{cosentino2018}, however, reported the presence of widespread and narrow SiO emission (average linewidth of 1.6 \kms) toward the northwest of G034 at an angular resolution $\sim$30\arcsec. 

Our ALMA images reveal that SiO in G034 is organised into two plane-parallel structures that seem to impact onto the north-western dense ridge of G034. In a way analogue to a sea wave plunging onto a shore, the molecular gas pushed away by the W44 SNR (the sea wave) encounters the denser material of the IRDC (the beach), plunges and breaks causing the formation of two plane-parallel wave fronts or MHD shocks seen in SiO. As shown in Section$\,$\ref{model}, such a kinematic structure is best reproduced by a non-stationary CJ-type MHD shock, where the J-type discontinuity is seen in the PV diagrams as an almost vertical velocity drop, and the C-type part shows a shallower deceleration over $\sim$10\arcsec. The different velocity shapes of the SiO North and South PV diagrams may reflect inhomogeneities in the gas and/or slightly different physical conditions of the pre-shock gas (the derived $\zeta$ differ by a factor of 2 between PV North and South). The proximity of W44 is responsible for the high $\zeta$ and B$_{mag}$ required by the models. 

The dynamical ages predicted for the shocks in PV North and PV South (t$_{dyn}$=16000-17000$\,$years) are in agreement with the age of the SNR ($\sim$20000$\,$years) and the SiO depletion time-scales \citep[$\sim$10$^4$-10$^5$$\,$years;][]{martinpintado1992}. We also caution that the MHD$\_$VODE code only provides 1D highly idealised models that do not consider projection effects, which may explain any disagreement between the model predictions and the observations (especially for PV South).
 
Finally, we propose that the shock interaction in G034 between the expanding molecular gas from W44 and the pre-existing IRDC, may have shaped the Av$\geq$20$\,$mag ridge into its filamentary morphology, since the latter appears as an almost detached structure from the IRDC and since the SiO emission follows the shape of the ridge. 

This is supported by the fact that the density of the post-shock component in G034 is a factor $>$10 higher than in the pre-shock gas (n(H$_2$)$>$10$^{5}$$\,$cm$^{-3}$; Section$\,$\ref{c18o}), which may enable star formation. The lack of any cold core in the ridge is consistent with the idea that we are observing the earliest stages of such a process, since the age of the shock ($<$20000$\,$years) is too short to see the effects of gravitational collapse of any newly formed high-density structure or core \citep[time-scales $\sim$10$^5$$\,$years for densities of n(H$_2$)$>$10$^{5}$$\,$cm$^{-3}$;][]{mckee2003}.

In summary, we report the first detection of a fully resolved, time-dependent CJ-type MHD shock. This shock has been produced by the interaction between a molecular flow expanding from the SNR W44 and a high-density ridge part of the nearby IRDC G034.77-00.55. The high magnetic field and cosmic-rays ionization rate induced by the SNR is responsible for the large dissipation region of the CJ-type shock. The post-shock gas is compressed by the shock to densities n(H)$>$10$^5$$\,$cm$^{-3}$, comparable to those required for the formation of massive stars.

\section{Acknowledgments}
\noindent
This paper makes use of the ALMA data:\\ ADS/JAO.ALMA$\,$\#2016.1.01363.S. ALMA is a partnership of ESO (representing its member states), NSF (USA) and NINS (Japan), together with NRC (Canada), MOST and ASIAA (Taiwan), and KASI (Republic of Korea), in cooperation with the Republic of Chile. The Joint ALMA Observatory is operated by ESO, AUI/NRAO and NAOJ. This paper also makes use of observations carried out with the IRAM 30m telescope under projects 025-12. IRAM is supported by INSU/CNRS (France), MPG (Germany), and IGN (Spain). G.C. acknowledges financial support from University College London Perren/Impact Studentship and from the ESO Studentship Program Europe 2017/2018. I.J.-S. acknowledges the support from the STFC through an Ernest Rutherford Fellowship (grant number ST/L004801), and from the MINECO and FEDER funding under grant ESP2017-86582-C4-1-R. A.T.B acknowledges funding from the European Union Horizon 2020 research and innovation programme (grant agreement No 726384).



\end{document}